\documentclass[aps,prl,preprint,groupedaddress,showpacs,superscriptaddress]{revtex4-1}

\usepackage{graphicx} \usepackage{amsmath} \usepackage{epsfig}

\begin{document}

%number in the upper righthand corner of the title page in preprint
%mode.  % Multiple \preprint commands are allowed.  % Use the
%'preprintnumbers' class option to override journal defaults % to
%display numbers if necessary %\preprint{}

\title{Non-Thermal Electron Energization from Magnetic Reconnection in
  Laser-Driven Plasmas}

\author{S. Totorica}  
\affiliation{Kavli Institute for Particle Astrophysics and Cosmology, Stanford University, SLAC National Accelerator Laboratory, Menlo Park, CA 94025, USA}
\author{T. Abel} 
\affiliation{Kavli Institute for Particle Astrophysics and Cosmology, Stanford University, SLAC National Accelerator Laboratory, Menlo Park, CA 94025, USA}
\author{F. Fiuza}  \email{fiuza@slac.stanford.edu} 
\affiliation{High Energy Density Science Division, SLAC National Accelerator Laboratory, Menlo Park, CA 94025, USA}

\date{\today}

\begin{abstract} The possibility of studying non-thermal electron
  energization in laser-driven plasma experiments of magnetic
  reconnection is studied using two- and three-dimensional
  particle-in-cell simulations.  It is demonstrated that non-thermal
  electrons with energies more than an order of magnitude larger than
  the initial thermal energy can be produced in plasma conditions
  currently accessible in the laboratory.  Electrons are accelerated
  by the reconnection electric field, being injected at varied
  distances from the X-points, and in some cases trapped in plasmoids,
  before escaping the finite-sized system.  Trapped electrons can be
  further energized by the electric field arising from the motion of
  the plasmoid.  This acceleration gives rise to a non-thermal
  electron component that resembles a power-law spectrum, containing
  up to $\sim 8\%$ of the initial energy of the interacting electrons and
  $\sim 24\%$ of the initial magnetic energy.  Estimates of the maximum
  electron energy and of the plasma conditions required to observe
  suprathermal electron acceleration are provided, paving the way for
  a new platform for the experimental study of particle acceleration
  induced by reconnection.
\end{abstract}



\maketitle

Magnetic reconnection is a fundamental plasma process whereby magnetic
energy is rapidly and efficiently converted into plasma flows,
heating, and potentially non-thermal particles \cite{Zweibel2009}.  It
plays a critical role in the evolution of magnetized plasmas in space
physics, astrophysics and laboratory nuclear fusion devices
\cite{Ji2011,Joglekar2014,Taylor1986}.  Energetic particles
are a common signature of the reconnection process, and reconnection
is thought to be a promising candidate for producing the non-thermal
emissions associated with explosive phenomena such as solar flares,
pulsar wind nebulae and jets from active galactic nuclei.  These
systems span a large range of plasma conditions, and many recent
studies have focused on understanding the details of the particle
acceleration in both nonrelativistic
\cite{Oka2010,Fu2006,Hoshino2001,Drake2006b,Dahlin2014} and relativistic
\cite{Guo2014,Sironi2014a} regimes.  These studies have discovered a
rich variety of acceleration mechanisms, with many mediated by the
plasmoids that can form in the current sheet due to the tearing
instability \cite{Loureiro2007}.  The overall efficiency of
reconnection in producing energetic particles and its dependence on
the plasma conditions, however, has not been settled, and thus remains
an important and active area of research.

Recently, laser-driven high-energy-density (HED) plasmas have started
to be used to study reconnection in the laboratory
\cite{Nilson2006a,Li2007a,Nilson2008a,Willingale2010a,Zhong2010a,Dong2012a,Fiksel2014a}
(for a review of laboratory reconnection experiments see \cite{Yamada2010}).
By focusing terawatt class ($\sim$kJ/ns) lasers onto solid foils,
plasma bubbles are produced that expand and generate megagauss scale
toroidal magnetic fields by the Biermann battery ($\nabla n \times
\nabla T$) mechanism \cite{Haines1997}.  The expansion of two bubbles
placed in close proximity can then drive reconnection between the
self-generated or externally imposed \cite{Fiksel2014a} magnetic
fields (Figure 1(a)).  Many of the prominent features of reconnection
have been observed in these systems, including plasma jets, plasma
heating, changes in magnetic field topology, and plasmoid formation.
Supersonic and super-Alfv\'{e}nic inflow speeds place these experiments in
a regime of strongly-driven reconnection, making them particularly
relevant for systems featuring colliding magnetized plasma flows found
in space and astrophysics.  Laser-driven plasma experiments are
characterized by a high Lundquist number and large system size
relative to the electron and ion inertial lengths, allowing a
comparison with astrophysical systems via scaling laws
\cite{Ryutov2000}.  Until now, the possibility of using laser-driven
plasmas to study non-thermal particle acceleration induced by
reconnection has remained unclear.

In this Letter, we investigate the onset and the properties of
particle acceleration during reconnection in laser-driven plasmas
using \textit{ab initio} particle-in-cell (PIC) simulations.  Using
the fully relativistic, state-of-the-art PIC code OSIRIS
\cite{Hemker2015,Fonseca2002a,Fonseca2008,Fonseca2013} we perform
two-dimensional (2D) and, for the first time, three-dimensional (3D)
simulations in realistic experimental conditions and geometries.  We
demonstrate the possibility of detecting electrons with energies more
than an order of magnitude larger than the thermal energy in
conditions relevant to current experiments.  The electrons are
accelerated primarily by the reconnection electric field, with the randomness
associated with injection at varied distances from the X-points and
escape from the finite-sized system leading to a non-thermal component
with a power-law shape.  Furthermore, a fraction of the electrons can
be trapped in plasmoids, slowly gaining further energy due to the
drifting motion of the plasmoid.  An estimate for the maximum electron
energy and a threshold condition for suprathermal energization are
given in terms of experimentally tunable parameters, which can
guide future studies of particle acceleration induced by
reconnection in laser-driven plasmas.

We simulate the interaction between two expanding, magnetized plasma
bubbles, consistent with previous PIC studies
\cite{Fox2011a,Fox2012a,Lu2013a,Lu2014} (Figs. 1(b) and 1(c)).  The
simulations start when the bubbles are about to interact, and thus do
not model the initial generation of the plasma or magnetic field.
However, they capture the most important features of the system such
as finite size and driven inflows, allowing them to be connected to a
number of experimental geometries \cite{Ryutov2013a}.  The bubbles are
centered at ${\bf R}^{(1)} = (0, R, 0)$ and ${\bf R}^{(2)} = (0, -R,
0)$, with the radial vectors from the bubble centers being ${\bf
  r}^{(i)} = {\bf r} - {\bf R}^{(i)}$.  Here $R$ is the radius of each
bubble when they begin to interact.  The initial density profile is
$n_{b} + n^{(1)} + n^{(2)}$ where $n^{(i)}(r^{(i)}) = (n_{0} -
n_{b})\,\textup{cos}^{2}\left ( \frac{\pi r^{(i)}}{2 R_{0}} \right )$
if $r^{(i)} < R_{0}$, 0 otherwise.  Here, $n_{b} = 0.01 n_{0}$ is a
background plasma density, and $R_{0}$ is the initial bubble radius,
which we typically take to be $0.9 - 1.0 R$.  The velocity profile is
${\bf V}^{(1)} + {\bf V}^{(2)}$ where $ {\bf V}^{(i)}(r^{(i)}) =
V_{0}\,\textup{sin}\left ( \frac{\pi r^{(i)}}{R_{0}} \right ) {\bf
  r}^{(i)}$ if $r^{(i)} < R_{0}$, 0 otherwise.  The initial magnetic
field is divergence free and corresponds to the sum of two oppositely
aligned ribbons, ${\bf B}^{(1)} + {\bf B}^{(2)}$ where $ {\bf
  B}^{(i)}(r^{(i)}) = B_{0} f(\theta^{(i)}) \,\textup{sin} \left (
  \frac{\pi (R_{0} - r^{(i)})}{2 L_{B}} \right ) {\bf
  \hat{\phi}}^{(i)}$ if $R_{0} - 2 L_{B} \leq r^{(i)} \leq R_{0}$, 0
otherwise.  $L_{B} = R_{0} / 4$ is the initial half-width of the
magnetic field ribbon.  $\theta^{(i)}$ and $\phi^{(i)}$ are the polar
and azimuthal angles, respectively, of spherical coordinate systems
originating at the bubble centers.  For the 2D simulations, which
model the plane $\theta^{(i)} = \pi / 2$, $f(\theta^{(i)}) = 1$.  For
the 3D simulations, $f(\theta^{(i)})$ is zero near the z-axis, rises
from 0 to 1 from $\theta^{(i)} = \pi / 16$ to $\theta^{(i)} = 2 \pi /
16$, and similarly from $\theta^{(i)} = 15 \pi / 16$ to $\theta^{(i)}
= 14 \pi / 16$.  The rise is given by the polynomial $10 x^{3} - 15
x^{4} + 6 x^{5}$ which approximates a Gaussian shape.  The motional
electric field is initialized as ${\bf E} = - {\bf V} \times {\bf B} /
c$, where ${\bf V} = {\bf V}_{e} = {\bf V}_{i}$ at $t=0$.  The 2D (3D)
simulations use a box of size $8 R \times 2 R$ ($4 R \times 2 R \times
4 R$) and are evolved to $t / t_{d} = 0.5 (0.25)$, where $t_{d} = R /
V_{0}$ is the relevant timescale for the interaction.  The boundaries
along the outflow directions are absorbing for the particles initially
in the plasma bubble, thermal re-emitting for the particles initially
in the background plasma, and an absorbing layer \cite{Vay2000a} for
the electromagnetic fields.  The energy spectra presented are
integrated over the electrons initially in the bubbles.  Simulations
with larger domains confirm that the boundaries are not affecting the
results.  Periodic boundaries are used along the inflow direction,
taking advantage of the antisymmetry of the system.  We note that the
majority of reconnection simulation studies employ periodic boundaries
in the direction of the reconnection outflows and thus may suffer
unphysical effects from recirculating particles and radiation.
Studying particle acceleration in the finite-sized systems of
laser-driven plasmas would yield critical insight for understanding
the role of boundary conditions and particle escape in models of
reconnection.

To directly match laboratory conditions of interest (e.g. those
generated with the OMEGA EP laser \cite{Fiksel2014a}) we model a range
of Alfv\'{e}nic and sonic Mach numbers, $M_{A} = V_{0} / V_{A} = 4 -
64$ and $M_{S} = V_{0} / C_{S} = 2 - 8$, where $V_{A} = B_{0} /
\sqrt{4 \pi n_{0} m_{i}}$ is the Alfv\'{e}n speed and $C_{S} = \sqrt{Z
  T_{e} / m_{i}}$ is the sound speed.  We also match the high
experimental plasma beta, $\beta_{e} = \frac{n_{0} T_{e}}{B_{0}^{2} /
  8 \pi} = 2 \left ( \frac{M_{A}}{M_{S}} \right )^{2} = 8 - 128$.  The
bubble radii are $R / (c / \omega_{pi}) \simeq 26.5 \, (20)$ for the
2D (3D) simulations, within the experimentally accessible range of $R
/ (c / \omega_{pi}) = 10-100$ \cite{Fox2012a}.  The initial
temperatures $T_{e}$ and $T_{i}$ are taken to be equal and uniform.
The parameters $C_{S} / c$ and $m_{i} / m_{e} Z$ (where $c$ is the
speed of light, $m_{i}$ and $m_{e}$ the ion and electron masses, and
$Z$ the ion charge) are not directly matched to experimental values
due to computational expense.  In the majority of the simuations we
use $V_{0} / c = 0.1$ and $m_{i} / m_{e} Z = 128$, with $C_{S}$ chosen
to correctly match $M_{S}$.  By varying these parameters in the range
$V_{0} / c = 0.025 - 0.1$ and $m_{i} / m_{e} Z = 32 - 512$ we have
verified that the main results of this study are relatively
insensitive to their exact values.  All simulations use a spatial grid
of resolution of $\Delta x = 0.5 c / \omega_{pe} \simeq 0.04 c /
\omega_{pi}$ and cubic interpolation for the particles.  The 2D (3D)
simulations use a timestep of $\Delta t = 0.35 (0.285) \;
\omega_{pe}^{-1}$ and 64 (8) particles per cell per species.

We first analyze in detail the 2D simulation with $M_{S} = 2$ and
$M_{A} = 4$.  We observe that the early evolution of the system is
consistent with previous studies using this configuration
\cite{Fox2011a,Fox2012a,Lu2013a,Lu2014}.  As the plasma flows bring in
the magnetic flux at super-Alfv\'{e}nic speeds, the plasma is
compressed by the ram pressure and the amplitude of the magnetic field
increases by a factor of $\approx 1.65$.  A current sheet forms on the
order of the ion inertial length, which is the scale at which
electrons and ions decouple, enabling fast reconnection mediated by
the Hall effect \cite{Pritchett2001}.  The current sheet is then
unstable to the tearing instability, and we observe the formation of a
single plasmoid, consistent with linear theory which predicts $k_{max}
\delta = 0.55$, where $\delta \approx 1.41 \, c / \omega_{pi}$ is the
half-width of the current sheet \cite{Pritchett1991}.  For the range
of conditions simulated we typically observe magnetic field
enhancement by a factor of 1.5 to 5 and the formation of 1-3
plasmoids.  The calculation of the different terms in the generalized
Ohm's law (${\bf E} = - \frac{1}{c} \frac{n_{i}}{n_{e}} {\bf v}_{i}
\times {\bf B} + \frac{1}{e n_{e} c} {\bf J} \times {\bf
  B}-\frac{\nabla \cdot {\bf P}_{e}}{e n_{e}} - \frac{m_{e}}{e}\frac{d
  {\bf v}_{e}}{dt}$) shows that the electric field in the
diffusion region is predominantly supported by the Hall and
off-diagonal electron pressure tensor terms, in agreement with
previous kinetic studies of reconnection starting from the Harris
equilibrium \cite{Pritchett2001} and laser-driven reconnection
\cite{Fox2011a}.  For the geometry of our configuration the typical
value of the reconnection electric field in the diffusion region is $E
\approx 0.5 V_{0} B_{0}$ (Figure 1(c)).  In terms of the local
Alfv\'{e}n speed and compressed magnetic field, we find $E \approx 0.3
V_{A} B$ at $t / t_{d} = 0.25$, when the plasmoid is just beginning to
form.  After the onset of reconnection, the finite size of the plasma
bubbles results in reconnection outflows that are directed out of the
interaction region.

Figure 1(d) shows the temporal evolution of the electron energy
spectrum for this simulation.  We observe the development of a
non-thermal component with energies up to $\approx 50 \, k_{B} T_{e}$ that
resembles a power law with index $P \approx -5.3$.  In Figure 2(a), a
distribution of energetic electrons is seen in the reconnection
region.  The energy decreases with distance from x = 0, which is
approximately the location of the X-point before the plasmoid forms.
At a later time, shown in Figure 2(b), energetic electrons are seen
both to escape in the plasma outflows and to be trapped in a peak at
the location of the single plasmoid that forms in the simulation.  The
most energetic electrons are trapped at $t / t_{d} = 0.5$.  In a
realistic 3D system, which we will discuss later, the trapped
electrons would also escape the reconnection region, but along the
z-direction.

In order to understand the details of the particle acceleration, we
tracked the detailed motion of the most energetic electrons in the
simulation.  The relative importance of the electric field components
in energizing the electrons was determined by calculating the work
done by each throughout the simulation, $W_{i} = \int_{0}^{t} {dt}'
(p_{i} / \gamma \,m_{e}) (-e \, E_{i})$.  Figs. 2(c) and 2(d) show
trajectories that illustrate the two distinct types of behavior
observed for the energetic electrons.  The evolution of the total
energy and the work done by the electric field components for these
electrons is plotted in Figs. 2(e) and 2(f).  In both cases the
particles gain the majority of their energy as they are accelerated in
the out-of-plane direction by the reconnection electric field near the
X-point, as can be seen in the insets in Figs. 2(c) and 2(d).  The
in-plane magnetic field then rotates the velocities of the electrons
into the plane, at which point the electrons will either escape in the
outflow direction (as in Figure 2(c)), become trapped in a plasmoid
(as in Figure 2(d)), or travel around the bubble along the field
lines.  The electrons that escape the reconnection region lose some of
their energy to the in-plane polarization electric field (Figure
2(e)), whereas the trapped electrons can be further accelerated due to
the motion of the plasmoid.  The energization rate is seen to decrease
for the trapped electrons, as shown in Figure 2(f).  In this
simulation the plasmoid is slowly drifting along the outflow
direction, and thus the trapped electrons see an out-of-plane motional
electric field with alternating polarity (Figure 1(c)) as they circle
inside the plasmoid \cite{Oka2010}.  Plasma flowing into the drifting
plasmoid from the bounding X-points leads to an asymmetry in its
motional electric field, allowing trapped electrons to gain further
energy.  Simulations with a single bubble confirm that the electric
field from the bubble expansion does not significantly energize
particles.

In Figure 3(a) we observe that the low-energy portion of the electron
distribution is well fitted by a Maxwell-Boltzmann distribution (blue
dashed line) and that there is a non-thermal component that starts at
$\sim 5 k_{B} T_{e}$ and resembles a power-law with an index of $\sim
5.3$ and an exponential cutoff at $\sim 20 k_{B} T_{e}$ (red dashed
line). Periodic simulations for the same plasma parameters but
transversely infinite profiles (not shown here) show that in much
larger systems the power-law component extends to higher energies and
has a harder index of $\sim 3.5$, indicating the importance of
particle escape in finite systems. The comparison with transversely
infinite systems will be discussed in detail elsewhere. At the end of
the simulation ($t / t_{d} = 0.5$), the non-thermal component contains
$\sim 8\%$ of the initial energy of interacting electrons, $\sim 24\%$
of the initial energy stored in the magnetic field, and $\sim 1\%$ of
the total number of electrons initially in the bubbles.  Previous work
has shown that the presence of plasmoids may give rise to
Fermi-acceleration mechanisms \cite{Drake2006b}.  Calculating
$d\epsilon / dt$ as a function of $\epsilon$ for the tracked particles
(where $\epsilon$ is energy) shows an energization rate that is
approximately constant, indicating that the dominant acceleration
mechanism producing the energetic electrons is not Fermi acceleration.
This is likely to require a much larger number of plasmoids and thus a
more energetic laser drive than that used in current experiments.  We
note that even in the simulations with up to 3 plasmoids we do not
observe evidence of Fermi acceleration associated with the plasmoids.
In our case, the distribution of energies for the accelerated
electrons is established as electrons are injected into the diffusion
region at varied distances from the X-points and escape the diffusion
region at different times.  The electrons that interact nearer to the
X-point are exposed to the reconnection electric field for a longer
amount of time before the in-plane magnetic fields direct them out of
the diffusion region, allowing them to reach higher energies
\cite{Sironi2014a}.  Additional randomness is introduced by the finite
probability for a given electron to escape from the system or become
trapped in a plasmoid and further energized.

Figure 3(b) shows a comparison of the electron energy spectra at $t /
t_{d} = 0.25$ for four 2D simulations with $M_{S} = 2$ and $M_{A}$
ranging from 4 to 32.  All show the development of a high energy tail
with a power-law shape and similar spectral indices, indicating the
similarity of the acceleration mechanisms across these conditions.
The maximum electron energy attained increases with the initial
magnetic field amplitude, consistent with the picture that the
electrons gain their energy directly from the reconnection electric
field.  Figure 3(c) shows the evolution of the maximum electron energy
in each simulation, and the corresponding energization rates are shown
in Figure 3(d).  The effective electric field corresponding to the
energization rate is found to be close to $0.5 V_{0} B_{0}$ for all
cases, consistent with the typical value of the reconnection electric
field observed in the diffusion region in the simulations.

2D simulations do not capture the finite size of the system in the
out-of-plane (acceleration) direction.  It is thus critical to
consider 3D effects.  To this end we have performed full 3D
simulations of reconnection between expanding plasma bubbles, with the
initialization described above and the conditions $M_{S} = 2, M_{A} =
4$.  They show that the out-of-plane variation does not significantly
change the acceleration picture but does limit the maximum electron
energy attainable.  An important difference to the 2D case is the fact
that the amplitude of the magnetic field is only seen to increase by
$\approx 7\%$, in contrast with the $\approx 65\%$ increase in
2D.  The electron spectrum at $t / t_{d} = 0.25$ is shown in Figure
3(b).  The spectrum exhibits a high-energy tail with a similar
spectral index to the 2D case, demonstrating that the non-thermal
energization survives in 3D and that the acceleration mechanisms are
similar.  The maximum electron energy is smaller for the 3D case, due
to the finite out-of-plane size limiting the maximum energy
attainable.  The energization rate is also slightly smaller in 3D, due
to the fact that the electrons do not see a uniform value of the
reconnection electric field as they travel in the out-of-plane
direction.

The finite out-of-plane size limits the distance over which an
electron can be accelerated by the reconnection electric field to
approximately the diameter of the bubble.  Considering the effective
value of the reconnection electric field is $\approx 0.5 V_{0} B_{0}$,
an estimate for the maximum energy increase of an electron can then be
written in terms of experimentally tunable parameters as
$\epsilon_{max} / k_{B}T_{e} = \left ( M_{S}^2 / M_{A} \right ) \left(
  R_{b} / (c / \omega_{pi}) \right )$.  Applying this to the
parameters of the 3D simulation presented above gives $\epsilon_{max}
/ k_{B}T_{e} \approx 20$, which predicts well the location of the
exponential cutoff of the power-law as shown in Figure 3(b).  The
threshold condition for producing suprathermal electrons in the
laboratory is then $\left ( M_{S}^2 / M_{A} \right ) \left( R_{b} / (c
  / \omega_{pi}) \right ) > 1$, which is easily satisfied by the
conditions accessible in current experiments.  The conditions of two
recent laser-driven reconnection experiments at the Omega Laser
Facility \cite{Fox2012a,Fiksel2014a} are estimated to be $M_{S}
\approx 2.5,5.6$, $M_{A} \approx 20,9.2$, and $R_{b} / (c /
\omega_{pi}) \approx 80,22$.  For these conditions our model predicts
maximum electron energies of $\epsilon_{max} / k_{B}T_{e} \approx
25,75$.  An important experimental signature of this acceleration due
to reconnection is that the non-thermal electrons should be detected
in a fan-like profile, with energetic electrons being emitted both in
the direction of reconnection outflows and in the direction opposite
to the reconnection electric field.

In summary, by performing \textit{ab initio} kinetic simulations of
magnetic reconnection in laser-driven plasmas, we have shown that
these systems can accelerate non-thermal electrons to energies more
than an order of magnitude larger than the thermal energy in
conditions currently accessible in the laboratory.  Electron injection
at varied distances from the X-points and escape from the finite-sized
system leads to a non-thermal component that resembles a power-law
spectrum.  Our results indicate that laser-driven plasmas thus offer a
new platform for the experimental study of particle acceleration
induced by reconnection, which could help illuminate the role
reconnection plays in explosive phenomena associated with space and
astrophysical plasmas.

\begin{acknowledgements}
  This work was supported by the U.S. Department of Energy SLAC
  Contract No. DE-AC02-76SF00515. The authors acknowledge the OSIRIS
  Consortium, consisting of UCLA and IST (Portugal) for the use of the
  OSIRIS 3.0 framework and the visXD framework.  S. T. was supported
  by the NCSA through the Blue Waters Graduate Fellowship program and
  the Department of Defense through the NDSEG fellowship program.
  This work was also partially supported by DOE Fusion Energy Science,
  the SLAC LDRD program, and by the NSF Grant (PICKSC), AC11339893.
  Simulations were run on Mira (ALCF supported under Contract
  No. DE-AC02-06CH1135) through an INCITE award, on Blue Waters, and
  on the Bullet Cluster at SLAC.
\end{acknowledgements}

\bibliographystyle{apsrev4-1}

\newpage

\begin{figure}[h]
\begin{center}
\includegraphics[width=0.9\textwidth]{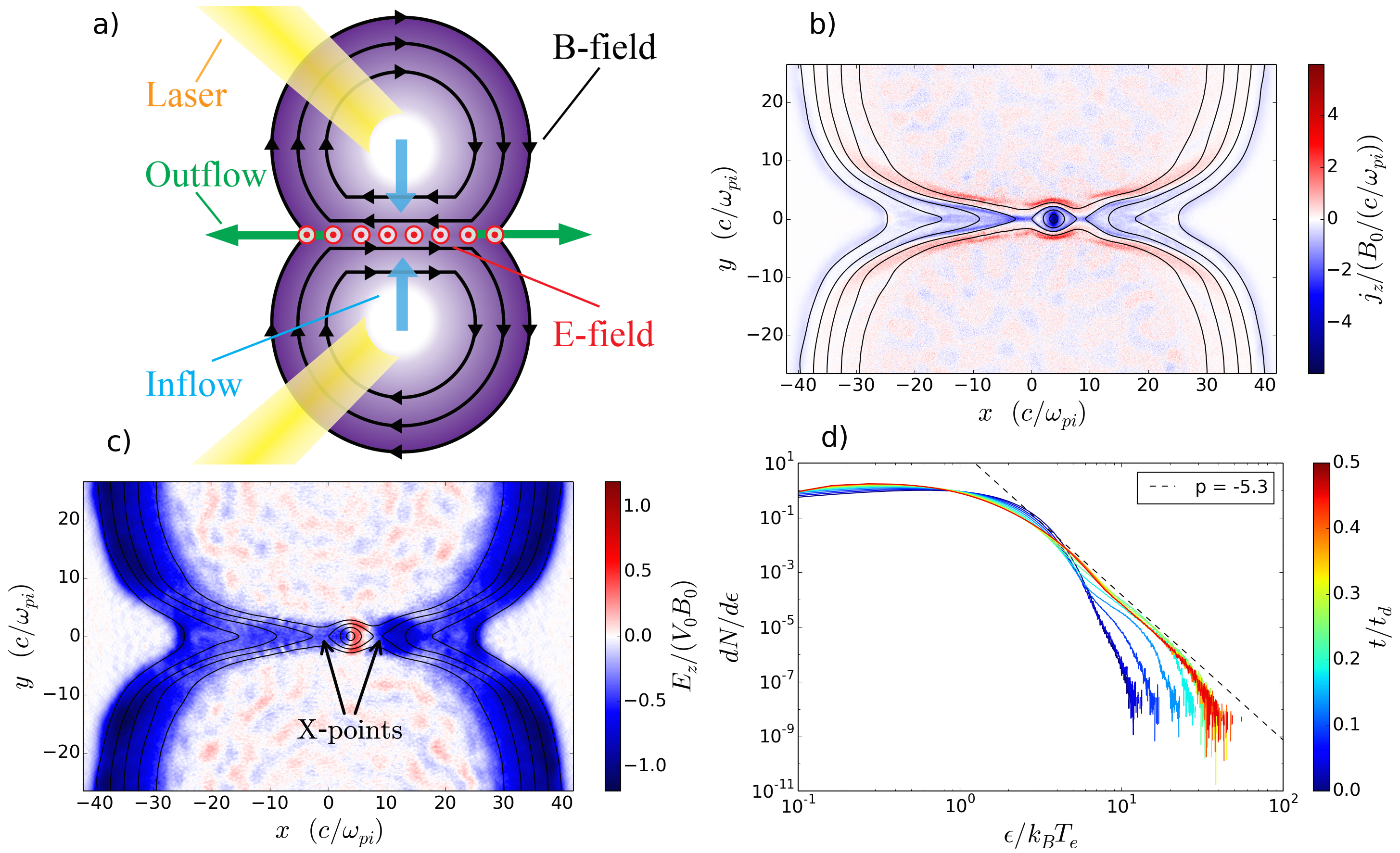}
\caption{\label{fig:geometry} a) Geometry of laser-driven reconnection
  experiments and simulations.  b) Out-of-plane current density and c)
  Out-of-plane electric field, with overlayed magnetic field lines.
  At the X-points, this is the reconnection electric field.  d)
  Evolution of the electron energy spectrum.}
\end{center}
\end{figure}

\begin{figure}[h]
\begin{center}
\includegraphics[width=0.9\textwidth]{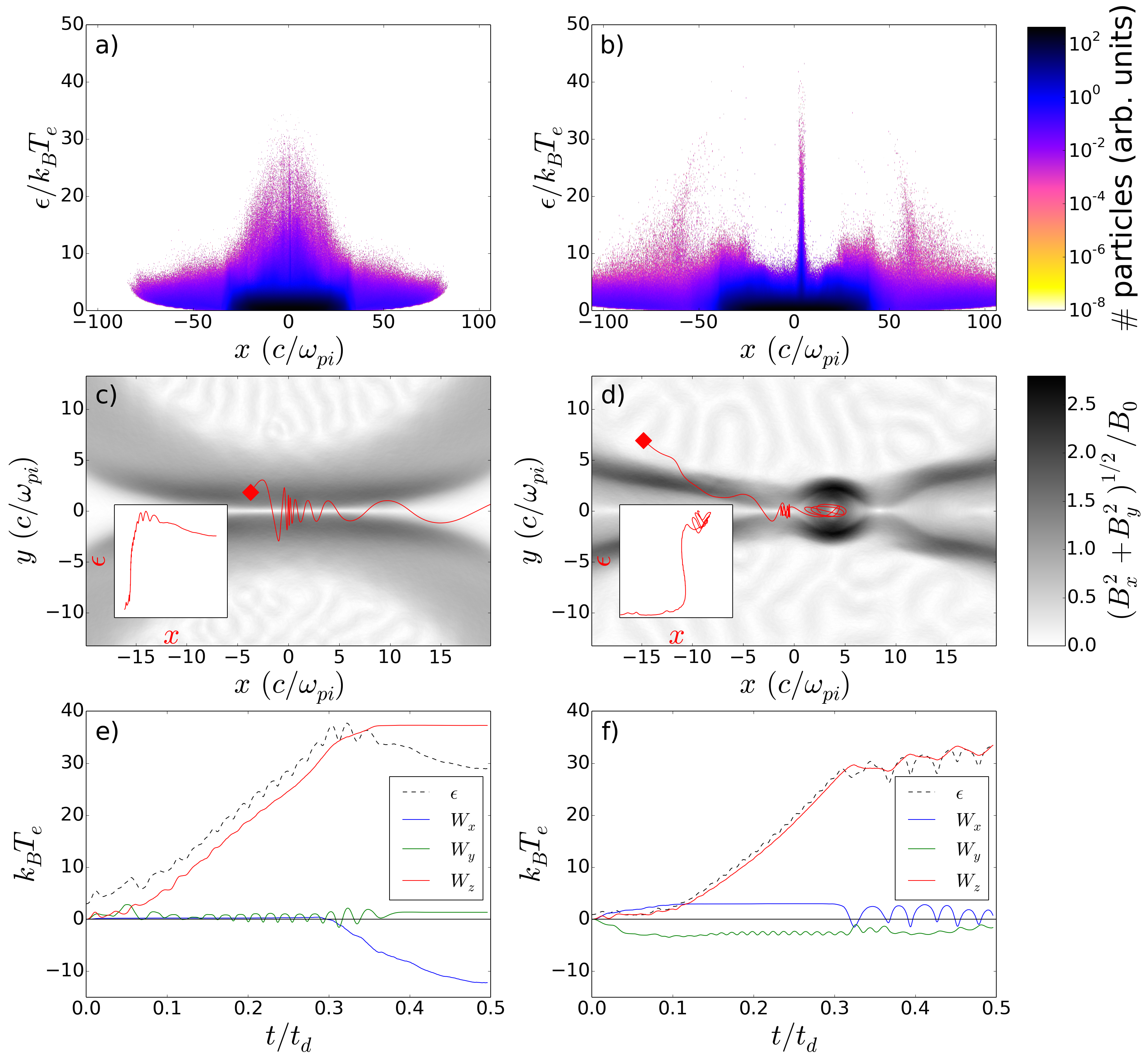}
\caption{\label{fig:2d_bubbles} a), b) Electron energy spectrum along
  the x-direction at $t / t_{d} = 0.25, 0.5$, respectively.  c)
  Trajectory of an escaping electron and d) Trajectory of a trapped
  electron, plotted over the magnitude of the in-plane magnetic field
  at $t / t_{d} = 0.25, 0.5$, respectively.  Diamonds indicate the
  particle's position at $t = 0$, and insets show the energy variation
  of the particle along x.  e), f) Evolution of the total energy and
  work done by the electric field components for the electrons shown
  in c), d), respectively.}
\end{center}
\end{figure}

\begin{figure}[h]
\begin{center}
\includegraphics[width=0.9\textwidth]{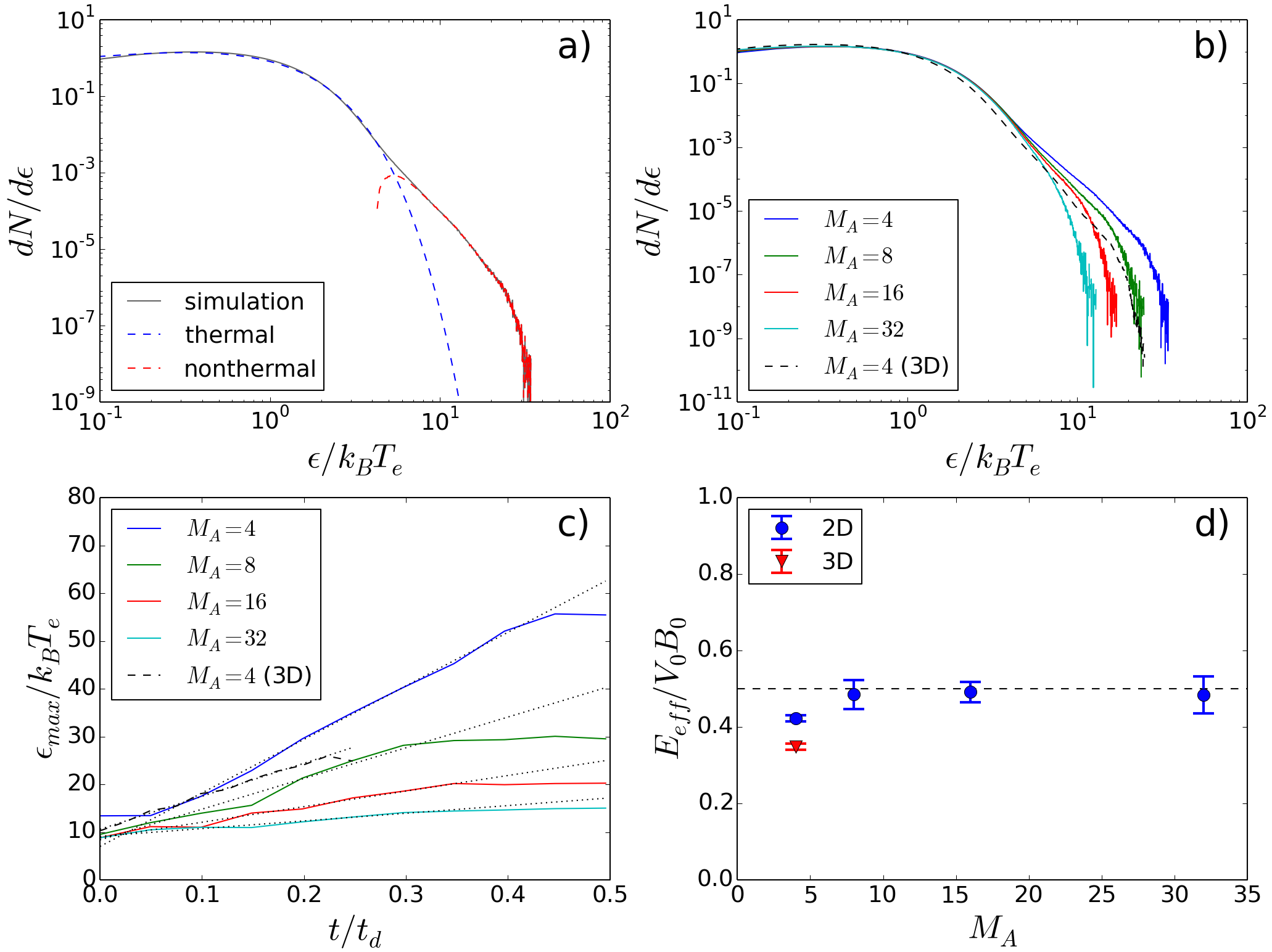}
\caption{\label{fig:tracking} a) Electron energy spectrum for the
  $M_{S} = 2$, $M_{A} = 4$ simulation at $t / t_{d} = 0.25$, with the
  thermal component (blue dashed line) and the non-thermal component
  (red dashed line).  b) Comparison of the electron energy spectra for
  different $M_{A}$ in 2D and 3D at $t / t_{d} = 0.25$.  c) Temporal
  evolution of the maximum electron energy for the same simulations as
  in b), with linear fits plotted as dotted lines.  d) Effective
  electric field obtained from the data in c).  Error bars show the
  one standard deviation error obtained from the linear fit.}
\end{center}
\end{figure}

\end{document}